# Method for Computation of the Fisher Information Matrix in the Expectation–Maximization Algorithm

**Lingyao Meng**

**Department of Applied Mathematics & Statistics**

**The Johns Hopkins University**

**Baltimore, Maryland 21218**

## Abstract

The expectation–maximization (EM) algorithm is an iterative computational method to calculate the maximum likelihood estimators (MLEs) from the sample data. It converts a complicated one-time calculation for the MLE of the incomplete data to a series of relatively simple calculations for the MLEs of the complete data. When the MLE is available, we naturally want the Fisher information matrix (FIM) of unknown parameters. The FIM is, in fact, a good measure of the amount of information a sample of data provides and can be used to determine the lower bound of the variance and the asymptotic variance of the estimators. However, one of the limitations of the EM is that the FIM is not an automatic by-product of the algorithm. In this paper, we review some basic ideas of the EM and the FIM. Then we construct a simple Monte Carlo-based method requiring only the gradient values of the function we obtain from the E step and basic operations. Finally, we conduct theoretical analysis and numerical examples to show the efficiency of our method. The key part of our method is to utilize the simultaneous perturbation stochastic approximation method to approximate the Hessian matrix from the gradient of the conditional expectation of the complete-data log-likelihood function.

*Key words*: Fisher information matrix, EM algorithm, Monte Carlo, Simultaneous perturbation stochastic approximation

# 1. Introduction

The expectation–maximization (EM) algorithm introduced by Dempster, Laird and Rubin in 1977 is a well-known method to compute the MLE iteratively from the observed data. It is applied to problems in which the observed data is incomplete or the log-likelihood function of the observed data is so complicated that obtaining the MLE from it directly is too difficult. There are some books on EM algorithm such as McLachlan and Krishnan (1997), Watanabe and Yamaguchi (2004), and Gupta and Chen (2011). The superiority of the EM algorithm comes from its simplicity. It obtains a sequence of results from the simple complete data log-likelihood function, thereby avoiding calculations from the complicated incomplete data log-likelihood function. Dempster, Laird, and Rubin have proved that this sequence has the monotonicity property of increase with each iteration. However, because the incomplete (i.e., observed) data log-likelihood function is not directly used in the EM algorithm, we cannot directly obtain the Fisher information matrix (FIM) for the observed data. This paper presents a general method for computing the FIM in the EM setting.

The FIM plays a key role in uncertainty calculation and other aspects of estimation for a wide range of statistical methods and applications, including parameter estimation (e.g., Abt and Welch, 1998; Spall, 2014), system identification (Ljung, 1999, pp. 212–221), and experimental design (Pronzato, 2008; Spall, 2010), among other areas. Hence, it is critical to obtain the FIM in a practical manner. Both the expected FIM (the expectation of the square of the gradient of the incomplete data log-likelihood functions) at the MLE and the observed FIM (the negative Hessian matrix of the incomplete data log-likelihood functions evaluated at the observed data) at the MLE can be used to calculate the Cramer-Rao lower bound and asymptotic distribution of the MLE. Some methods have been introduced to estimate the FIM. In Spall (2005), for example, the author constructed an efficient method to approximate the expected FIM based on Monte Carlo simulation. Extensions are given in Spall (2008) and Das et al. (2010). Moreover, Berisha and Hero (2015) introduced a method to estimate the FIM directly from the sample data without the density function.

Given the importance of the FIM and the popularity of the EM, some research has been done to calculate the FIM from the EM. Louis (1982) provides a closed form of the observed FIM (the negative Hessian matrix of incomplete data) that requires the negative conditional expectation of the Hessian matrix of the complete data with respect to the incomplete data and the conditional expectation of the square of gradient of complete data log-likelihood function with respect to the incomplete data. In Meng and Rubin (1991), the authors introduced the supplemented EM (SEM) algorithm to calculate the observed FIM.

However, all these methods have limitations. One common disadvantage is that we can only obtain the observed information matrix from these methods, yet the expected Fisher information matrix at MLE is a maximum likelihood estimator of the expected Fisher information matrix at the true value. Furthermore, Cao and Spall (2012) and Cao (2013) demonstrated that:

> *Under reasonable conditions similar to standard MLE conditions, the inverse expected FIM outperforms the inverse observed FIM under a mean squared error criterion.*

In addition, the formulas from Louis (1982) work well for some relatively simple problems. Nevertheless, when the log-likelihood function of the complete data is more complicated, the calculating of the negative conditional expectation of the Hessian matrix of the complete data log-likelihood function with respect to the incomplete data is difficult to accomplish. Additionally, the conditional expectation of the square of the complete data score function is likewise hard to calculate. In the SEM algorithm, similarly, the calculation of the negative conditional expectation of the Hessian matrix is sometimes infeasible. The SEM algorithm, furthermore, requires running the EM algorithm many times, and the number of iterations depends on the number of iterations in the EM algorithm. However, the cost of running the EM algorithm is high, and its convergence rate is slow.

Using simultaneous perturbation stochastic approximation (SPSA) (Spall, 1992, 1998; Bhatnagar et al., 2013), our method takes the numerical differentiation to the gradient of the conditional expectation of the complete-data log-likelihood function. If the gradient is unavailable, we can approximate it with the value of the conditional expectation of the complete-data log-likelihood function, which can be obtained by running the E step. Then through the Monte Carlo simulation, we can approximate the expected FIM by calculating the negative average of all the approximate Hessian matrices.

Section 2 in this article covers the basics of the FIM and the EM algorithm. Section 3 presents our Monte Carlo-based method. The numerical examples and theoretical analysis to show the efficiency of our method are in Sections 4 and 5, respectively.

## 2. Review of EM Algorithm

In order to have a good understanding for further discussion, we are going to introduce basic information on the EM algorithm, (the expected and the observed) FIM and some existing methods to calculate the FIM in the EM algorithm in this section.

### 2.1 The EM algorithm

Let $X$ and $Y$ be the missing data and the observed (incomplete) data, respectively; so $Z = (X, Y)$ is the complete data. We also suppose that the density function of $Z$ is:

$$p_Z(z|\theta) = p_{X,Y}(x, y|\theta),$$

and the density function of $Y$ is $p_Y(y|\theta)$. Let the log-likelihood function of the complete data be:

$$L(\theta|Z) = L(\theta|X, Y) = \log\{p_Z(z|\theta)\}.$$

Let the log-likelihood function of the observed data be:

$$L_O(\theta|Y) = \log\{p_Y(y|\theta)\}.$$

The purpose of the EM algorithm is to calculate the MLE of $\theta$, say $\theta^*$, from the observed data $Y$.

The EM algorithm operates as follows. The Expectation (E) step: Calculate the expectation of the log-likelihood function $L(\theta|X, Y)$ with respect to the conditional distribution of $X$ given $Y$ under the current estimate of the parameters $\theta^{(t)}$,

$$Q(\theta|\theta^{(t)}) = E[L(\theta|X, Y)|Y, \theta^{(t)}].$$

The Maximization (M) step: Find the parameters $\boldsymbol{\theta}$ that maximize $Q(\boldsymbol{\theta}|\boldsymbol{\theta}^{(t)})$ for fixed $\boldsymbol{\theta}^{(t)}$,

$$\boldsymbol{\theta}^{(t+1)} = \arg\max Q(\boldsymbol{\theta}|\boldsymbol{\theta}^{(t)}).$$

In each iteration of the EM algorithm, we need to run the E step and the M step only once. This process is repeated until the difference between $L_O(\boldsymbol{\theta}^{(t^*+1)}|Y)$ and $L_O(\boldsymbol{\theta}^{(t^*)}|Y)$ is less than a particular positive number $\delta$.

It was shown in Dempster, Laird and Rubin (1977) that the sequence $L_O(\boldsymbol{\theta}^{(t)}|Y)$ is a nondecreasing sequence. Further, it was shown in Wu (1983) and Vaida (2005) that the log-likelihood converges to the value at $\boldsymbol{\theta}^*$ and $\boldsymbol{\theta}^{(t)}$ converges to $\boldsymbol{\theta}^*$ when $t$ approaches infinity.

Define the mapping $\boldsymbol{M}(\boldsymbol{\theta}^{(t)}) = \boldsymbol{\theta}^{(t+1)}$ and $\boldsymbol{DM}$ is the Jacobian matrix of $\boldsymbol{M}$ at $\boldsymbol{\theta}^*$.

## 2.2 The Fisher Information Matrix

The FIM is a good measure of the amount of information the sample data can provide about parameters. Suppose $p(\boldsymbol{\theta};x)$ is the density function of the object model and $L(\boldsymbol{\theta};x) = \log(p(\boldsymbol{\theta};x))$ is the log-likelihood function. We can define the expected FIM as:

$$E\left[\frac{\partial L(\boldsymbol{\theta}|X)}{\partial \boldsymbol{\theta}} \frac{\partial L(\boldsymbol{\theta}|X)}{\partial \boldsymbol{\theta}^T}\right].$$

Under the condition in which we can interchange the order of integration and differentiation,

$$E\left[\frac{\partial L(\boldsymbol{\theta}|X)}{\partial \boldsymbol{\theta}}\right] = \int \frac{\partial L(\boldsymbol{\theta}|x)}{\partial \boldsymbol{\theta}} p(\boldsymbol{\theta};x)dx = \int \frac{\partial p(\boldsymbol{\theta}|x)}{\partial \boldsymbol{\theta}} dx = \frac{\partial}{\partial \boldsymbol{\theta}} \int p(\boldsymbol{\theta}|x)\, dx = 0. \qquad (2.2.1)$$

Therefore the expected FIM also equals:

$$\text{cov}\left[\frac{\partial L(\boldsymbol{\theta}|\boldsymbol{X})}{\partial \boldsymbol{\theta}}\right].$$

Additionally, from (2.2.1):

$$E\left[\frac{\partial L(\boldsymbol{\theta}|\boldsymbol{X})}{\partial \boldsymbol{\theta}}\frac{\partial L(\boldsymbol{\theta}|\boldsymbol{X})}{\partial \boldsymbol{\theta}^T}\right] = -E\left[\left(\frac{\partial^2 L(\boldsymbol{\theta}|\boldsymbol{X})}{\partial \boldsymbol{\theta}\partial \boldsymbol{\theta}^T}\right)\right] = -E[H(\boldsymbol{\theta}|\boldsymbol{X})]. \qquad (2.2.2)$$

We then define the observed FIM as:

$$\frac{\partial^2 L(\boldsymbol{\theta}|\boldsymbol{X})}{\partial \boldsymbol{\theta}\partial \boldsymbol{\theta}^T} = -\left(\frac{\partial^2 L(\boldsymbol{\theta}|\boldsymbol{X})}{\partial \boldsymbol{\theta}\partial \boldsymbol{\theta}^T}\right) = -H(\boldsymbol{\theta}|\boldsymbol{X}).$$

## 2.3  Some existing methods

Louis (1982) presented the following formula:

$$\boldsymbol{I}_{\text{Louis}}(\boldsymbol{\theta}^*) = -\left\{E\left[\left(\frac{\partial^2 L(\boldsymbol{\theta}|\boldsymbol{X},\boldsymbol{Y})}{\partial \boldsymbol{\theta}\partial \boldsymbol{\theta}^T}\right)|\boldsymbol{Y},\boldsymbol{\theta}^*\right]\right\}_{\boldsymbol{\theta}=\boldsymbol{\theta}^*}$$

$$-\left\{E\left[\left(\frac{\partial L(\boldsymbol{\theta}|\boldsymbol{X},\boldsymbol{Y})}{\partial \boldsymbol{\theta}}\frac{\partial L(\boldsymbol{\theta}|\boldsymbol{X},\boldsymbol{Y})}{\partial \boldsymbol{\theta}^T}\right)|\boldsymbol{Y},\boldsymbol{\theta}^*\right]\right\}_{\boldsymbol{\theta}=\boldsymbol{\theta}^*}$$

$$+\left\{E\left[\frac{\partial L(\boldsymbol{\theta}|\boldsymbol{X},\boldsymbol{Y})}{\partial \boldsymbol{\theta}}|\boldsymbol{Y},\boldsymbol{\theta}^*\right]E\left[\frac{\partial L(\boldsymbol{\theta}|\boldsymbol{X},\boldsymbol{Y})}{\partial \boldsymbol{\theta}^T}|\boldsymbol{Y},\boldsymbol{\theta}^*\right]\right\}_{\boldsymbol{\theta}=\boldsymbol{\theta}^*}.$$

The formula in Oakes (1999) is:

$$\boldsymbol{I}_{\text{Oakes}}(\boldsymbol{\theta}^*) = \left\{\frac{\partial^2 Q(\boldsymbol{\theta}|\boldsymbol{\theta}^*)}{\partial \boldsymbol{\theta}^2} + \frac{\partial^2 Q(\boldsymbol{\theta}|\boldsymbol{\theta}^*)}{\partial \boldsymbol{\theta}\partial \boldsymbol{\theta}^*}\right\}_{\boldsymbol{\theta}=\boldsymbol{\theta}^*},$$

whereas in the SEM algorithm,

$$\boldsymbol{I}_{\text{SEM}}(\boldsymbol{\theta}^*) = (\boldsymbol{I} - \boldsymbol{DM})\left\{E\left[-\left(\frac{\partial^2 L(\boldsymbol{\theta}|\boldsymbol{X},\boldsymbol{Y})}{\partial \boldsymbol{\theta}\partial \boldsymbol{\theta}^T}\right)|\boldsymbol{Y},\boldsymbol{\theta}^*\right]\right\}_{\boldsymbol{\theta}=\boldsymbol{\theta}^*},$$

where $I$ is the identity matrix.

## 3. Computation of the Expected FIM in the EM Algorithm

In this section, we continue to use the same notations as in Section 2.1. According to the laws of conditional probability,

$$p_{X,Y}(x,y|\theta) = p_{X|Y}(x|y,\theta)p_Y(y|\theta).$$

Then

$$\log\{p_{X,Y}(x,y|\theta)\} = \log\{p_{X|Y}(x|y,\theta)\} + \log\{p_Y(y|\theta)\}.$$

Take expectations on both sides of the above equation with respect to the conditional distribution of $X|Y, \theta^{(t)}$, and we have:

$$Q(\theta|\theta^{(t)}) = E[\log\{p_{X|Y}(x|y,\theta)\}|Y,\theta^{(t)}] + L_O(\theta|Y). \qquad (3.1)$$

If we take the derivative of $\theta$ and let $\theta = \theta^{(t)}$, under the conditions where we can interchange the order of expectation with respect to $X|Y$ and the differentiation of $\theta$, the equation (3.1) implies:

$$\left[\frac{\partial Q(\theta|\theta^{(t)})}{\partial \theta}\right]_{\theta=\theta^{(t)}} = E\left[\left[\frac{\partial p_{X|Y}(x|y,\theta)}{\partial \theta}\frac{1}{p_{X|Y}(x|y,\theta)}\right]_{\theta=\theta^{(t)}}\bigg|Y,\theta=\theta^{(t)}\right] + \left[\frac{\partial L_O(\theta|Y)}{\partial \theta}\right]_{\theta=\theta^{(t)}}. \qquad (3.2)$$

The first part of the right hand side of the equation (3.2) equals 0. Therefore,

$$\left[\frac{\partial Q(\theta|\theta^{(t)})}{\partial \theta}\right]_{\theta=\theta^{(t)}} = \left[\frac{\partial L_O(\theta|Y)}{\partial \theta}\right]_{\theta=\theta^{(t)}}. \qquad (3.3)$$

While we know nothing about the value of $L_O(\theta|Y)$ or, subsequently, the gradient of $L_O(\theta|Y)$, we can obtain the value of $Q(\theta|\theta^{(t)})$ by running the E step and then the gradient of $Q(\theta|\theta^{(t)})$. Note that both sides of the equation (3.3) can be regarded as functions of $\theta^{(t)}$. Therefore, we define:

$$S(\theta^{(t)}|Y) = \left[\frac{\partial Q(\theta|\theta^{(t)})}{\partial \theta}\right]_{\theta=\theta^{(t)}} \quad (3.4)$$

Equation (3.3) implies that we can use $S(\theta|Y)$ to take place of the gradient of $L_O(\theta|Y)$. Then the Hessian matrix can be obtained by numerical differentiation on $S(\theta|Y)$.

There are many numerical differentiation methods, but, due to its efficiency in multivariate problems, we prefer the SPSA method introduced by Spall (1992), Spall (2000), and Spall (2005). Let $\widehat{H}^{(k)}(\theta|Y_k)$ be the estimator of the Hessian matrix for the data set $Y_k$ and $G(\theta|Y_k)$ be the gradient of the log-likelihood function for the data set $Y_k$. The formula to approximate the Hessian matrix is:

$$\widehat{H}^{(k)}(\theta|Y_k) = \frac{1}{2}\left\{\frac{\delta G_k}{2}[\Delta_{k1}^{-1}, \Delta_{k2}^{-1}, \ldots \Delta_{kd}^{-1}] + \left(\frac{\delta G_k}{2}[\Delta_{k1}^{-1}, \Delta_{k2}^{-1}, \ldots \Delta_{kd}^{-1}]\right)^T\right\}, \quad (3.5)$$

where $\delta G_k \equiv G(\theta + \Delta_k|Y_k) - G(\theta - \Delta_k|Y_k)$ and $\Delta_k \equiv [\Delta_{k1}, \Delta_{k2}, \ldots \Delta_{kd}]^T$ is a random perturbation vector. According to Spall (2005), for any $k$ and any $1 \leq j \leq d$, $\Delta_{kj}$ should satisfy:

- $\Delta_{kj}$ is a mean-zero and a symmetrically distributed random variable;
- every $\Delta_{kj}$ is uniformly bounded and $E[|1/\Delta_{kj}|] < \infty$; and
- The $\Delta_{kj}$ are independent and identically distributed across $k$ and $j$.

A good choice that satisfies the above conditions is the Bernoulli distribution with half the probability being $c$ and half the probability being $-c$. Note in the right hand side of (3.5), we add the transpose term in order to make the estimator symmetric. Moreover, by combining (3.4) and (3.5), the formula to estimate the Hessian matrix of the incomplete data log-likelihood function at the MLE $\boldsymbol{\theta}^*$ for the data set $Y_k$ is:

$$\widehat{H}^{(k)}(\boldsymbol{\theta}^*|Y_k) = \frac{1}{2}\left\{\frac{\delta S_k}{2}\left[\Delta_{k1}^{-1}, \Delta_{k2}^{-1} \ldots \ldots \Delta_{kd}^{-1}\right] + \left(\frac{\delta S_k}{2}\left[\Delta_{k1}^{-1}, \Delta_{k2}^{-1} \ldots \ldots \Delta_{kd}^{-1}\right]\right)^T\right\}, (3.6)$$

where $\delta S_k = S(\boldsymbol{\theta}^* + \boldsymbol{\Delta}_k|Y_k) - S(\boldsymbol{\theta}^* - \boldsymbol{\Delta}_k|Y_k)$. In the case where the value of $S(\boldsymbol{\theta}^* \pm \boldsymbol{\Delta}_k|Y_k)$ is unavailable, we can approximate it by the numerical differentiation on $Q(\boldsymbol{\theta}|\boldsymbol{\theta}^* \pm \boldsymbol{\Delta}_k)$, utilizing the following formula and then substituting the estimator of $S(\boldsymbol{\theta}^* \pm \boldsymbol{\Delta}_k|Y_k)$ into (3.5).

$$\widehat{S}(\boldsymbol{\theta}^* \pm \boldsymbol{\Delta}_k|Y_k) = \frac{Q(\boldsymbol{\theta}^* \pm \boldsymbol{\Delta}_k + \widehat{\boldsymbol{\Delta}}_k|\boldsymbol{\theta}^* \pm \boldsymbol{\Delta}_k) - Q(\boldsymbol{\theta}^* \pm \boldsymbol{\Delta}_k - \widehat{\boldsymbol{\Delta}}_k|\boldsymbol{\theta}^* \pm \boldsymbol{\Delta}_k)}{2}\begin{pmatrix}\widehat{\Delta}_{k1}^{-1}\\ \widehat{\Delta}_{k2}^{-1}\\ \vdots\\ \widehat{\Delta}_{kd}^{-1}\end{pmatrix}. (3.7)$$

In equation (3.7), $\widehat{\boldsymbol{\Delta}}_k = [\widehat{\Delta}_{k1}, \widehat{\Delta}_{k2} \ldots, \widehat{\Delta}_{kd}]^T$ is a perturbation vector that satisfies the same conditions as $\boldsymbol{\Delta}_k$. The superiority of the SPSA lies in the fact that for each Hessian matrix estimation, the SPSA requires two values of $S$ when $S$ is available with respect to *2d* values of $S$ using the traditional finite difference method. When $S$ is unavailable, compared to $2d(d+1)$ values of $Q$ used in the finite difference method, SPSA only needs four $Q$ values for any dimension *d*.

From (2.2.2), we can approximate the expected FIM by calculating the negative average of all estimations of the Hessian matrix. The algorithm to compute the expected FIM operates as follows:

1. Run the EM algorithm until the difference between $L_O(\theta^{(t^*+1)}|Y)$ and $L_O(\theta^{(t^*)}|Y)$ is less than a particular $\delta$, and then obtain the MLE $\theta^*$, at which we would like to determine the Hessian matrix. Set the small number c in Bernoulli $\pm c$ distribution to generate the perturbation vectors $\Delta_k$ and $\widehat{\Delta}_k$ (if needed). Set the number $N$, which is the number of data sets we would like to generate.

2. Generate $N$ data sets by Monte Carlo simulation and denote them by $Y_1, Y_2, \ldots$ and $Y_N$ respectively.

3. Do one of the following two actions:

    - In the case where the values of $S$ for different $\theta$ and data sets are calculable, generate $N$ perturbation vectors following Bernoulli $\pm c$ distribution and denote them by $\Delta_1, \Delta_2$ and $\Delta_N$ respectively. Set $k = 1$. Jump to step 4.

    - If the values of $S$ are unavailable, generate $2N$ perturbation vectors following Bernoulli $\pm c$ distribution and denote them by $\Delta_1, \Delta_2, \ldots \Delta_N, \widehat{\Delta}_1, \widehat{\Delta}_2, \ldots$ and $\widehat{\Delta}_N$ respectively. Set $k = 1$. Jump to step 5.

4. Apply formula (3.6) to calculate $\widehat{H}^{(k)}(\theta^*|Y_k)$ using $\Delta_k$ and $Y_k$. Set $k = k + 1$. Jump to step 6.

5. Run the E step of the EM algorithm four times to get $Q(\theta^* \pm \Delta_k + \widehat{\Delta}_k|\theta^* \pm \Delta_k)$ and $Q(\theta^* \pm \Delta_k - \widehat{\Delta}_k|\theta^* \pm \Delta_k)$. Apply formula (3.7) to calculate $\widehat{S}(\theta^* \pm \Delta_k|Y_k)$. Substitute $\widehat{S}(\theta^* \pm \Delta_k|Y_k)$ into (3.6) to obtain $\widehat{H}^{(k)}(\theta^*|Y_k)$. Set $k = k + 1$. Jump to step 6.

6. Repeat step 4 or 5 until $k = N$. Finally, calculate the negative average of all $\widehat{H}^{(k)}(\theta^*|Y_k)$s, which is the approximation of the expected FIM.

We suppose the $\boldsymbol{\theta}^*$ is available after running the EM algorithm. To compute the expected information matrix, the cost of the above algorithm is $N$ data sets, $N$ perturbation vectors, $2N$ $\boldsymbol{S}$ values, and $O(N)$ basic operations if the $\boldsymbol{S}$ values are available. If not, our algorithm requires $N$ data sets, $2N$ perturbation vectors, $4N$ times E step, and $O(N)$ basic operations.

This algorithm also can apply to the observed information matrix. Specifically, we approximate the Hessian matrix for one data set at $\boldsymbol{\theta}^*$ plus/minus different perturbation vectors, and then calculate the negative average of them.

## 4. Theoretical Analysis

There are two parts in this session. One is the comparison between our algorithm and the SEM algorithm in Meng and Rubin (1991). The other part compares our algorithm with the SP-based method in Spall (2005) to approximate the FIM in general settings.

### 4.1 Comparison with the SEM algorithm

One important question is: Is our method better than previous methods to calculate the FIM in the EM algorithm? However, it seems impossible to compare our method to all the existing methods; therefore, weconsider the relative performance between our method and the famous SEM method.

Recall the SEM method. To approximate $\boldsymbol{DM}$, Meng and Rubin have:

$$r_{ij}^{(t)} = \frac{M_j\left(\theta_1^*, \theta_2^*, \ldots, \theta_i^{(t)}, \theta_{i+1}^* \ldots \theta_d^*\right) - M_j(\theta^*)}{\theta_i^{(t)} - \theta_i^*},$$

where $\boldsymbol{\theta}^* = (\theta_1^*, \theta_2^*, \ldots, \theta_i^* \ldots \theta_d^*)^T$ is the MLE acquired from the EM algorithm, and $\boldsymbol{\theta}^{(t)} = (\theta_1^{(t)}, \theta_2^{(t)} \ldots, \theta_i^{(t)} \ldots \theta_d^{(t)})^T$ is one point in the parameter space not equal to the $\boldsymbol{\theta}^*$ in any element.

Then Meng and Rubin obtained $r_{ij}$, the $ij$ element of $\boldsymbol{DM}$ when the sequence $r_{ij}^{(t^*)}, r_{ij}^{(t^*+1)} \ldots$ is stable for some $t^*$.

Intuitively, the SEM method may have three limitations:

- Although we can calculate the complete-data information matrix in some simple examples, such as the Gaussian mixture model and the multinomial model, sometimes the complete-data information matrix is not feasible in practice.
- From Meng and Rubin (1991), SEM requires roughly $(d+1)/2$ times as much computational time as the EM itself. However, the convergence rate of EM is slow, and the cost is high if it is in high dimensional parameter spaces.
- We can only obtain the observed information matrix from SEM. Although shown by Efron and Hinkley (1978) that the observed information matrix is better than the expected one, it only holds for one-parameter translation families and with conditioning on an ancillary statistic. Moreover in Cao and Spall (2012) and Cao (2013), it was shown that under reasonable conditions similar to standard MLE conditions, the inverse expected FIM outperforms the inverse observed FIM under a mean squared error criterion.

In the case where computation of the second derivative of $Q(\theta|\theta^{(t)})$ is not feasible, we cannot utilize the SEM algorithm. As a result, here we would like to compare the SEM algorithm with our SP-based algorithm when the complete data information matrix is available. In that the SEM is used to calculate the observed FIM, we are going to compare the two algorithms in the observed FIM calculation. That is, we would like to compare two algorithms; one approximates $\boldsymbol{DM}$ by the SEM, and the other one applies the SPSA to approximate $\boldsymbol{DM}$ with different perturbation vectors and calculates the average of them. Note that in the following analysis, we will ignore the cost of basic operations, as the iterations of the EM algorithm are usually the dominant expense.

The formula to calculate the estimation of the $ij$th element of $\boldsymbol{DM}$ with $k$th perturbation vector from SPSA is:

$$\frac{M_j(\theta^* + \Delta_k) - M_j(\theta^* - \Delta_k)}{2\Delta_{ki}}. \tag{4.1.1}$$

Using Taylor Expansion and taking expectation to (4.1) we obtain:

$$E\left[\sum_{s=1}^{d} \frac{(\boldsymbol{DM})_{sj}\Delta_{ks}}{\Delta_{ki}}\right] + O(c^2). \tag{4.1.2}$$

Suppose $\Delta_{ks}$ follows Bernoulli $\pm c$ distribution, and because the $\Delta_{ks}$ are independent, $E[\Delta_{ks}/\Delta_{ki}] = 1$ when $s = i$ and $E[\Delta_{ks}/\Delta_{ki}] = 0$ when $s \neq i$. Therefore, (4.2) equals:

$$(\boldsymbol{DM})_{ij} + O(c^2).$$

In as much as the $c$ here can be arbitrarily small, we can ignore this term. Due to the independence of every element in $\Delta_k$, for $s \neq t$, $\text{cov}(\Delta_{ks}/\Delta_{ki}, \Delta_{kt}/\Delta_{ki}) = 0$. The variance of $\Delta_{ks}/\Delta_{ki}$ is 1 for $s \neq j$ and 0 for $s = j$. Therefore, the variance of (4.1) is $\sum_{s \neq i}[(\boldsymbol{DM})_{is}]^2$.

By the Central Limit Theorem, when we calculate the average of the SPSA term, we can obtain:

$$\frac{\sum_{k=1}^{n}\left(\sum_{s=1}^{d}\frac{(\boldsymbol{DM})_{sj}\Delta_{ks}}{\Delta_{ki}}\right)}{n} \sim N\left((\boldsymbol{DM})_{ij}, \frac{\sum_{s \neq i}[(\boldsymbol{DM})_{is}]^2)}{n}\right).$$

As $(\sum_{s \neq i}[(\boldsymbol{DM})_{is}(\boldsymbol{\theta}^*)]^2) \leq (d-1)(M_i^1)^2$, where $M_i^1$ is $\max(|DM|)_{ij}$, for $1 \leq j \leq d$ and $j \neq i$. If we let $n = d - 1$, by the 68–95–99.7 rule, the probability of the event, which is the average we obtain that lies in $(\boldsymbol{DM})_{ij} \pm 2M_i^1$, is at least 95%. When $n = d - 1$, we need $2(d-1)$ values of $M$ function, which requires $2(d-1)$ iterations of the EM algorithm.

Next we are going to consider the accuracy of the SEM algorithm when we run $2d$ times EM algorithm. Unfortunately, it is almost impossible to measure it accurately because the cost and accuracy of the SEM depends on the number of iterations of the EM, the starting value of the EM, the starting point of the SEM, and the fraction of missing information. However, we can value it under some reasonable assumptions. In Meng and Rubin (1991), the choice of the starting point of the SEM is the tradeoff between accuracy and computational cost. Meng and Rubin recommended using the second iterate of EM algorithm. Here, we follow their suggestions and make the following assumptions:

- The $\boldsymbol{M}_j$ is second differentiable, supposing that the $ii$th element of second derivative of $\boldsymbol{M}_j$ at $\boldsymbol{\theta}^*$ is $M_i'$.

- The starting value of the EM algorithm makes $|\theta_i^{(3)} - \theta_i^*| \geq \frac{4M_i^1}{|M_i'|}$ for every $1 \leq i \leq d$ hold, where $\theta_i^{(3)}$ is the $i$th element of $\boldsymbol{\theta}^{(3)}$. Under these assumptions, within the requirements of cost ($2d$ times EM algorithm), applying SEM algorithm, we can obtain:

$$r_{ij}{}^{(2)} = r_{ij} = \frac{\boldsymbol{M}_j\left(\boldsymbol{\theta}^* \pm \frac{4M_i^1}{|M_i'|}\boldsymbol{\delta}_i\right) - \boldsymbol{M}_j(\boldsymbol{\theta}^*)}{\frac{4M_i^1}{|M_i'|}} \approx (\boldsymbol{DM})_{ij} \pm 2M_i^1.$$

We have shown that the probability of the event, which is the approximation we obtain from the SPSA using $2(d-1)$ times EM algorithm that lies in $(\boldsymbol{DM})_{ij}(\boldsymbol{\theta}^*) \pm 2M_i^1$, is at least 95%. That is to say, under approximately equal cost ($2(d-1)$ for the SPSA versus $2d$ for the SEM), we are 95% sure that the $DM$ provided by the SPSA is more accurate.

## 4.2  Comparison with SP-based method in Spall (2005)

The SP-based method in Spall (2005) is also a feasible algorithm to approximate the FIM in EM. In order to approximate the expected (or observed) FIM, while we take the gradient of $Q(\boldsymbol{\theta}|\boldsymbol{\theta}^*)$ in our algorithm using SPSA, the method in Spall (2005) approximate the gradient of the incomplete data log-likelihood function. (Note that enhanced versions of the basic method in Spall, 2005, are considered in Spall, 2008, and Das et al., 2010; we do not consider these enhancements here.)

In this part, we assume the function value ($Q$ in our algorithm versus $L_O$ in Spall (2005)) is the only cost in these algorithms. Furthermore, we suppose one $Q$ value and one $L_O$ share the same cost although the $L_O$ is likely more complicated than $Q$. Otherwise, it may not necessary to apply EM algorithm.

From (3.1), we obtain:

$$L_O(\boldsymbol{\theta}|Y) = Q(\boldsymbol{\theta}|\boldsymbol{\theta}^*) - E\big[\log\{p_{X|Y}(x|y,\boldsymbol{\theta})\}\big|Y,\boldsymbol{\theta}^*\big]. \quad (4.2.1)$$

Let us define:

$$C(\boldsymbol{\theta}|\boldsymbol{\theta}^*) = E\big[\log\{p_{X|Y}(x|y,\boldsymbol{\theta})\}\big|Y,\boldsymbol{\theta}^*\big].$$

According to the property of EM algorithm, we have:

$$\left[\frac{\partial Q(\boldsymbol{\theta}|\boldsymbol{\theta}^*)}{\partial \boldsymbol{\theta}}\right]_{\boldsymbol{\theta}=\boldsymbol{\theta}^*} = \left[\frac{\partial L_O(\boldsymbol{\theta}|Y)}{\partial \boldsymbol{\theta}}\right]_{\boldsymbol{\theta}=\boldsymbol{\theta}^*} = \left[\frac{\partial C(\boldsymbol{\theta}|\boldsymbol{\theta}^*)}{\partial \boldsymbol{\theta}}\right]_{\boldsymbol{\theta}=\boldsymbol{\theta}^*} = \mathbf{0}. \quad (4.2.2)$$

Based on (4.2.1):

$$\frac{L_O(\boldsymbol{\theta}^* + \boldsymbol{\Delta}_k|Y) - L_O(\boldsymbol{\theta}^* - \boldsymbol{\Delta}_k|Y)}{2}\begin{pmatrix}\Delta_{k1}^{-1}\\ \Delta_{k2}^{-1}\\ \vdots\\ \Delta_{kd}^{-1}\end{pmatrix}$$

$$= \frac{Q(\boldsymbol{\theta}^* + \boldsymbol{\Delta}_k|\boldsymbol{\theta}^*) - Q(\boldsymbol{\theta}^* - \boldsymbol{\Delta}_k|\boldsymbol{\theta}^*)}{2}\begin{pmatrix}\Delta_{k1}^{-1}\\ \Delta_{k2}^{-1}\\ \vdots\\ \Delta_{kd}^{-1}\end{pmatrix}$$

$$- \frac{C(\boldsymbol{\theta}^* + \boldsymbol{\Delta}_k|\boldsymbol{\theta}^*) - C(\boldsymbol{\theta}^* - \boldsymbol{\Delta}_k|\boldsymbol{\theta}^*)}{2}\begin{pmatrix}\Delta_{k1}^{-1}\\ \Delta_{k2}^{-1}\\ \vdots\\ \Delta_{kd}^{-1}\end{pmatrix}. \quad (4.2.3)$$

With the assumption that $L_O, Q$, and $C$ are several times differentiable at $\boldsymbol{\theta}^*$, take expectation to (4.2.3) and combine it with (4.2.2), the bias of the expectation of *m*th component of the gradient approximation using the general SP-based method is:

$$b_{km} = \frac{1}{12} E\left[\frac{L'''_o(\overline{\theta}^{*(+)}) - L'''_o(\overline{\theta}^{*(-)})}{\Delta_{km}}[\Delta_k \otimes \Delta_k \otimes \Delta_k]\right]$$

$$= \frac{1}{12} E\left[\frac{\left(\frac{\partial^3 Q(\overline{\theta}^{*(+)}|\theta^*)}{\partial \theta^3} - \frac{\partial^3 C(\overline{\theta}^{*(+)}|\theta^*)}{\partial \theta^3}\right) + \left(\frac{\partial^3 Q(\overline{\theta}^{*(-)}|\theta^*)}{\partial \theta^3} - \frac{\partial^3 C(\overline{\theta}^{*(-)}|\theta^*)}{\partial \theta^3}\right)}{\Delta_{km}}[\Delta_k \otimes \Delta_k \otimes \Delta_k]\right],$$

(4.2.4)

where $\overline{\theta}^{*(\pm)}$ denotes pointes between $\theta^*$ and $\theta^* \pm \Delta_k$ and $\otimes$ is the Kronecker product.

The bias from our algorithm is:

$$b_{km}' = \frac{1}{12} E\left[\frac{\frac{\partial^3 Q(\overline{\theta}^{*(+)}|\theta^*)}{\partial \theta^3} + \frac{\partial^3 Q(\overline{\theta}^{*(-)}|\theta^*)}{\partial \theta^3}}{\Delta_{km}}[\Delta_k \otimes \Delta_k \otimes \Delta_k]\right]. \qquad (4.2.5)$$

Given that the $c$ can be small, (4.2.4) and (4.2.5) imply that both methods provide almost unbiased estimator of the gradient, which is 0 in this setting, and the difference between their biases are also ignorable.

## 5. Numerical Examples

In this section, we introduce two numerical examples, one for the mixture of two Gaussian distributions from Louis (1982) and the other for parameters in a state-space model. In the first example, to calculate the expected FIM and the observed FIM, we use our algorithm; in addition, we also utilize Louis' formula on the observed FIM. In the state-space model, we apply our algorithm to obtain the expected FIM.

## 5.1 Mixture of Gaussian Distributions

Let $Y_1, Y_2 \ldots, Y_n$ be the i.i.d. observed data from the Gaussian mixture model:

$$p_Y(y|\pi, \mu_1, \mu_2) = (1-\pi)\phi(y-\mu_1) + \pi\phi(y-\mu_2)$$

with unknown parameter $\boldsymbol{\theta} = \{\pi, \mu_1, \mu_2\}$ satisfying:

$$0 \leq \pi \leq 1, \quad -\infty < \mu_1, \mu_2 < +\infty.$$

$\phi(y)$ is the density function of standard normal distribution. We would like to calculate the expected and observed FIM at $\boldsymbol{\theta}^*$.

Let $Z_i = (X_i, Y_i)$ $i = 1, \ldots, n$ be the complete data where $X_1, \ldots, X_n$ represents the missing data satisfying:

$$X_i = \begin{cases} 0 & \text{when } Y_i \text{ is from } N(\mu_1, 1) \\ 1 & \text{when } Y_i \text{ is from } N(\mu_2, 1) \end{cases}.$$

Therefore, the density function of the complete data is:

$$p_{X,Y}(x, y|\pi, \mu_1, \mu_2) = (1-x)(1-\pi)\phi(y-\mu_1) + x\pi\phi(y-\mu_2).$$

It is reasonable to apply the EM algorithm to this problem with the observed data $Y_i$, the missing data $X_i$, and the complete data $Z_i$. The log-likelihood function of the observed data, $L_O(\boldsymbol{\theta}|Y)$, is:

$$\log p_Y(y_i|\pi, \mu_1, \mu_2) = \log\bigl((1-\pi)\phi(y_i - \mu_1) + \pi\phi(y_i - \mu_2)\bigr).$$

The log-likelihood function of the complete data, $L(\boldsymbol{\theta}|Z) = L(\boldsymbol{\theta}|X, Y)$, is:

$$(1 - x_i)\log(1 - \pi) + x_i \log \pi + (1 - x_i)\log\phi(y_i - \mu_1) + x_i \log\phi(y_i - \mu_2).$$

Let the conditional expectation of $X_i$ given $Y_i$ and $\boldsymbol{\theta}^{(t)} = \left(\pi^{(t)}, \mu_1^{(t)}, \mu_2^{(t)}\right)^T$ be $\alpha_i^{(t)}$:

$$\alpha_i^{(t)} = E[X_i|Y_i, \boldsymbol{\theta}^{(t)}] = \frac{\pi^{(t)}\phi(Y_i - \mu_2^{(t)})}{(1-\pi^{(t)})\phi(Y_i - \mu_1^{(t)}) + \pi^{(t)}\phi(Y_i - \mu_2^{(t)})}. \tag{5.1.1}$$

From Louis (1982), the EM operator $\boldsymbol{M}(\boldsymbol{\theta}^{(t)})$ is:

$$\boldsymbol{M}(\boldsymbol{\theta}^{(t)}) = \boldsymbol{\theta}^{(t+1)} = \begin{pmatrix} \pi^{(t+1)} \\ \mu_1^{(t+1)} \\ \mu_2^{(t+1)} \end{pmatrix} = \begin{pmatrix} \sum_{i=1}^n \alpha_i^{(t)}/n \\ \sum_{i=1}^n (1-\alpha_i^{(t)})Y_i / (n - \sum_{i=1}^n \alpha_i^{(t)}) \\ \sum_{i=1}^n \alpha_i^{(t)} Y_i / \sum_{i=1}^n \alpha_i^{(t)} \end{pmatrix}.$$

Set the true $\boldsymbol{\theta} = (\pi, \mu_1, \mu_2)^T = (2/3, 3, 0)^T$ and generate 750 data from the following density function:

$$p_Y\left(y \mid \tfrac{2}{3}, 0, 3\right) = \left(1 - \tfrac{2}{3}\right)\phi(y - 3) + \tfrac{2}{3}\phi(y).$$

Set the starting value $\boldsymbol{\theta}^{(0)} = (0.1, 1.0, 1.0)^T$. We obtain $\boldsymbol{\theta}^* = (0.6625, 3.0133, -0.0148)^T$ using EM algorithm after 54 iterations.

To apply our algorithm to compute the expected FIM, firstly, we should know the form of $\boldsymbol{S}(\boldsymbol{\theta}'|Y_i)$ for different $\boldsymbol{\theta}' = (\theta_1', \theta_2', \theta_3')^T$:

$$\boldsymbol{S}(\boldsymbol{\theta}'|Y_i) = \left[\frac{\partial Q(\boldsymbol{\theta}|\boldsymbol{\theta}')}{\partial \boldsymbol{\theta}}\right]_{\boldsymbol{\theta}=\boldsymbol{\theta}'} = \begin{pmatrix} E[X_i|Y_i, \boldsymbol{\theta}']/\theta_1' - (1 - E[X_i|Y_i, \boldsymbol{\theta}'])/(1 - \theta_1') \\ (1 - E[X_i|Y_i, \boldsymbol{\theta}'])(Y_i - \theta_2') \\ E[X_i|Y_i, \boldsymbol{\theta}'](Y_i - \theta_3') \end{pmatrix} \tag{5.1.2}$$

We can calculate $\boldsymbol{S}(\boldsymbol{\theta}'|Y_i)$ for different $\boldsymbol{\theta}'$ and $Y_i$ by substituting (5.1.1) into (5.1.2). Then we generate pseudodata $Y_i$ from the mixture Gaussian distribution based on $\boldsymbol{\theta}^*$.

The expected FIM obtained from our algorithm is:

$$\begin{pmatrix} 2680.7 & -180.9 & -266.8 \\ -190.9 & 171.4 & -65.5 \\ -266.8 & -65.6 & 396.0 \end{pmatrix}.$$

To compute the observed FIM by Louis' method, it is required to calculate the complete-data information matrix, which is:

$$\begin{pmatrix} \frac{X_i}{\pi^2} + \frac{1-X_i}{(1-\pi)^2} & 0 & 0 \\ 0 & 1-X_i & 0 \\ 0 & 0 & X_i \end{pmatrix} = \begin{pmatrix} 3354.1 & 0 & 0 \\ 0 & 253.2 & 0 \\ 0 & 0 & 496.8 \end{pmatrix},$$

the conditional expectation of the square of gradient of complete data log-likelihood function and (5.1.2) at $\boldsymbol{\theta}^*$. After calculation, the observed FIM from Louis' method, which is the same as the true observed FIM is:

$$I_{\text{Louis}}(\boldsymbol{\theta}^*) = \begin{pmatrix} 2708.3 & -200.1 & -237.1 \\ -200.1 & 176.6 & -59.0 \\ -237.1 & -59.0 & 395.2 \end{pmatrix}.$$

Based on our algorithm, the estimator of the observed FIM at $\boldsymbol{\theta}^*$ is:

$$I_{\text{SPSA}}(\boldsymbol{\theta}^*) = \begin{pmatrix} 2709.5 & -205.0 & -235.8 \\ -205.0 & 178.7 & -61.9 \\ -235.8 & -61.9 & 396.5 \end{pmatrix}.$$

We can measure the accuracy of our method by calculating $||I_{\text{Louis}}(\boldsymbol{\theta}^*) - I_{\text{SPSA}}(\boldsymbol{\theta}^*)||/||I_{\text{Louis}}(\boldsymbol{\theta}^*)||$, where $||\cdot||$ is the spectral norm.

$$\frac{||I_{\text{Louis}}(\boldsymbol{\theta}^*) - I_{\text{SPSA}}(\boldsymbol{\theta}^*)||}{||I_{\text{Louis}}(\boldsymbol{\theta}^*)||} = 0.0029.$$

$I_{\text{Louis}}(\boldsymbol{\theta}^*)$ is the true value of the observed FIM evaluated at $\boldsymbol{\theta}^*$. The result shows that the observed FIM obtained from our SPSA-based method is a good approximation to the true observed FIM evaluated at $\boldsymbol{\theta}^*$.

## 5.2  State-space model

A state-space model is a model that use state variables to represent a physical system. It is defined by the two equations, state equation (5.2.1) and measurement equation (5.2.2).

$$\boldsymbol{x}_t = \boldsymbol{A}\boldsymbol{x}_{t-1} + \boldsymbol{w}_t, \qquad (5.2.1)$$

$$\boldsymbol{y}_t = \boldsymbol{D}\boldsymbol{x}_t + \boldsymbol{v}_t, \qquad (5.2.2)$$

for $t = 1, \dots, n$. In (5.2.1) and (5.2.2), while $\boldsymbol{x}_t$ is an unobserved $p$-dimensonal state process, we can observe $\boldsymbol{y}_t$, which is a $q$-dimensonal process. $\boldsymbol{A}$ is a $p \times p$ transition matrix and $\boldsymbol{D}$ is a $q \times q$ measurement matrix. Noises $\boldsymbol{w}_t$ and $\boldsymbol{v}_t$ are typically assumed independent, Gaussian distributed, with means $\boldsymbol{0}$ and covariance matrices $\text{cov}(\boldsymbol{w}_t) = \boldsymbol{Q}$ and $\text{cov}(\boldsymbol{v}_t) = \boldsymbol{R}$. We also assume that the mean of $\boldsymbol{x}_0$, $\boldsymbol{\mu}$, and the covariance matrix of $\boldsymbol{x}_0$, $\boldsymbol{\Sigma}$, are known and $\boldsymbol{x}_0$, $\boldsymbol{w}_t$ and $\boldsymbol{v}_t$ are mutually independent. Note that the analytical form of the FIM for arbitrary parameters in a state-space model is difficult to obtain, but the analytical form *is* available in special cases, as illustrated in Spall and Garner (1990) and Cao (2013, Sect. 3.3).

In this model, $\boldsymbol{x}_0, \boldsymbol{x}_1 \dots, \boldsymbol{x}_n, \boldsymbol{y}_1, \dots \boldsymbol{y}_n$ is the complete data. However, we only have the observed data, which is $\boldsymbol{x}_0, \boldsymbol{y}_1, \dots \boldsymbol{y}_n$.

From Shumway and Stoffer (1982), neglecting the constant term, the log-likelihood function of the complete data in the state-space model above is:

$$L(\mu, \Sigma, A, Q, R | x_0, x_1 \ldots, x_n, y_1, \ldots y_n)$$

$$= -\frac{1}{2}\log|\Sigma| - \frac{1}{2}(x_0 - \mu)^T \Sigma^{-1}(x_0 - \mu)$$

$$-\frac{n}{2}\log|Q| - \frac{1}{2}\sum_{t=1}^{n}(x_t - A x_{t-1})^T Q^{-1}(x_t - A x_{t-1}) - \frac{n}{2}\log|R|$$

$$-\frac{1}{2}\sum_{t=1}^{n}(y_t - Dx_t)^T R^{-1}(y_t - Dx_t),$$

where $|*|$ means the determinant of "$*$".

In this section, we would like to consider the same setting in Cao (2013). Let $p = 3$ and $q = 1$. Suppose further that $\mu, \Sigma, A, D$, and $R$ are known and

$$\mu = (0 \quad 0 \quad 0)^T,$$

$$\Sigma = \begin{pmatrix} 0 & 0 & 0 \\ 0 & 0 & 0 \\ 0 & 0 & 0 \end{pmatrix},$$

$$A = \begin{pmatrix} 0 & 1 & 0 \\ 0 & 0 & 1 \\ 0.8 & 0.8 & -0.8 \end{pmatrix},$$

$$D = (1 \quad 0 \quad 0),$$

and $R = 1$.

Let $Q$ be a $3 \times 3$ diagonal matrix with diagonal elements $Q_1, Q_2,$ and $Q_3$. We are going to estimate $\theta = (Q_1, Q_2, Q_3)^T$ with true $\theta = (1,1,1)^T$. Applying the procedure in Shumway and Stoffer (1982) and setting $n = 100$, we obtain $\boldsymbol{\theta}^* = (0.9372, 0.9863, 1.0536)^T$ after 75 EM iterations.

After we know the $\boldsymbol{\theta}^*$, we can use our algorithm to approximate the expected Fisher information matrix at $\boldsymbol{\theta}^*$. In this study, given that the conditional expectation of complete data log-likelihood function is complicated, we apply (3.7) to obtain the gradient of the conditional expectation and then (3.6) to calculate the Hessian matrix.

When $N = 10000$,

$$I(\boldsymbol{\theta}^*) = \begin{pmatrix} 5.0430 & 2.3190 & 3.1588 \\ 2.3190 & 3.0455 & 1.5283 \\ 3.1588 & 1.5283 & 3.9970 \end{pmatrix},$$

$$\text{and} \quad \boldsymbol{I}_n(\boldsymbol{\theta}^*)^{-1} = \left(\frac{I(\boldsymbol{\theta}^*)}{n}\right)^{-1} = \begin{pmatrix} 48.9167 & -22.0781 & -30.3024 \\ -22.0781 & 50.3417 & -1.9311 \\ -30.3024 & -1.9311 & 49.6131 \end{pmatrix}.$$

When $N = 20000$,

$$I(\boldsymbol{\theta}^*) = \begin{pmatrix} 6.5415 & 3.3503 & 4.1475 \\ 3.3503 & 3.8149 & 2.2182 \\ 4.1475 & 2.2182 & 4.5081 \end{pmatrix},$$

$$\text{and} \quad \boldsymbol{I}_n(\boldsymbol{\theta}^*)^{-1} = \left(\frac{I(\boldsymbol{\theta}^*)}{n}\right)^{-1} = \begin{pmatrix} 47.7088 & -22.9402 & -32.6053 \\ -22.9402 & 47.7483 & -2.3891 \\ -32.6053 & -2.3891 & 53.3555 \end{pmatrix}.$$

Because there is no closed form for $\boldsymbol{I}_n(\boldsymbol{\theta}^*)^{-1}$, we apply the approximation in Cao (2013) as the true value for comparison. The result in Cao (2013) is:

$$\boldsymbol{I}_n(\boldsymbol{\theta}^*)^{-1}{}_{\text{true}} = \begin{pmatrix} 51.8934 & -24.4471 & -33.7404 \\ -24.4471 & 59.4544 & -3.36401 \\ -33.7404 & -3.36401 & 63.0565 \end{pmatrix}.$$

When $N = 10000$,

$$\frac{\left\|I_n(\boldsymbol{\theta}^*)^{-1} - I_n(\boldsymbol{\theta}^*)^{-1}{}_{\text{true}}\right\|}{\left\|I_n(\boldsymbol{\theta}^*)^{-1}{}_{\text{true}}\right\|} = 0.1499.$$

When $N = 20000$,

$$\frac{\left\|I_n(\boldsymbol{\theta}^*)^{-1} - I_n(\boldsymbol{\theta}^*)^{-1}{}_{\text{true}}\right\|}{\left\|I_n(\boldsymbol{\theta}^*)^{-1}{}_{\text{true}}\right\|} = 0.1259.$$

The result shows that the accuracy increases with $N$ increases.

## 6. Conclusions

Calculating the expected/observed Fisher information matrix (FIM) is often not easy in practice. It is even more difficult in the EM algorithm in that the EM algorithm obtained the result from a simple complete data log-likelihood function, not the incomplete data log-likelihood function associated with observed data. This article introduces a Monte Carlo-based method to approximate the expected FIM in the EM algorithm. We generate the pseudodata at $\boldsymbol{\theta}^*$ by the bootstrap method and then apply the numerical differentiation of the gradient of the function obtained from the E step for different pseudodata, acquiring the negative average. Compared to existing methods, for example the approach in Louis (1982) and the SEM in Meng and Rubin (1991), our method does not require the complete-data information matrix, which is sometimes complicated. We only require the values of $S(\boldsymbol{\theta}|Y)$ for different $\boldsymbol{\theta}$ or the values of $Q(\boldsymbol{\theta}|\boldsymbol{\theta}')$ for different $\boldsymbol{\theta}$. Theoretical analysis as well as numerical analyses are conducted to show the superiority of our method.